\documentclass[12pt,preprint]{aastex}
\newcommand{\be}{\begin{equation}}
\newcommand{\ee}{\end{equation}}
\newcommand{\bea}{\begin{eqnarray}}
\newcommand{\eea}{\end{eqnarray}}
\slugcomment{Submitted to Astrophysical Journal}
\shorttitle{Extrasolar Planets in Resonance}
\shortauthors{C. Beaug\'e, S. Ferraz-Mello and T.A. Michtchenko}
\begin{document}
\title{Extrasolar Planets in Mean-Motion Resonance: Apses Alignment and 
       Asymmetric Stationary Solutions}
\author{C. Beaug\'e\altaffilmark{1}}
\affil{Observatorio Astron\'omico, Universidad Nacional de C\'ordoba, Laprida
       854, (5000) C\'ordoba, Argentina}
\and
\author{S. Ferraz-Mello and T.A. Michtchenko}
\affil{Instituto Astron\^omico, Geof\'{\i}sico e Ci\^encias Atmosf\'ericas,
       Universidade de S\~ao Paulo, Rua do Mat\~ao 1226, 05508-900 S\~ao Paulo,
       Brasil}
\altaffiltext{1}{Present address: Instituto Nacional de Pesquisas Espaciais 
(INPE), Av. dos Astronautas 1758, (12227-010) S\~ao Jos\'e dos Campos, SP, 
Brasil}

\begin{abstract}
In recent years several pairs of extrasolar planets have been discovered 
in the vicinity of mean-motion commensurabilities. In some cases, such as the 
{\it Gliese 876} system, the planets seem to be trapped in a stationary 
solution, the system exhibiting a simultaneous libration of the resonant angle 
$\theta_1=2\lambda_2-\lambda_1-\varpi_1$ and of the relative position of the  
pericenters. 

In this paper we analyze the existence and location of these stable solutions, 
for the 2/1 and 3/1 resonances, as function of the masses and orbital elements 
of both planets. This is undertaken via an analytical model for the resonant 
Hamiltonian function. The results are compared with those of numerical 
simulations of the exact equations.

In the 2/1 commensurability, we show the existence of three principal families 
of stationary solutions: {\it (i)} aligned orbits, in which $\theta_1$ and 
$\varpi_1 - \varpi_2$ both librate around zero, {\it (ii)} anti-aligned orbits,
in which  $\theta_1=0$ and the difference in pericenter is $180$ degrees, and 
{\it (iii)} asymmetric stationary solutions, where both the resonant angle and 
$\varpi_1 - \varpi_2$ are constants with values different of 0 or 180 degrees.
Each family exists in a different domain of values of the mass ratio and 
eccentricities of both planets. Similar results are also found in the 3/1 
resonance.

We discuss the application of these results to the extrasolar planetary 
systems and develop a chart of possible planetary orbits with apsidal 
corotation. We estimate, also, the maximum planetary masses in order that the 
stationary solutions are dynamically stable.

\end{abstract}  


\section{Introduction}
As the number of known extrasolar planets continues to grow, so does the
discovery of multiple systems in which more than one planet seems to orbit the 
star. As of October 2002, ten planetary systems are known around main sequence 
stars, each with two or three members. 

From the point of view of Celestial Mechanics, the most interesting cases
are those in which two of the planets have orbital periods which are 
commensurable. Four extrasolar systems seem to satisfy this condition:
{\it Gliese 876}, {\it HD 82943}, {\it 55 Cnc} and {\it 47 Uma}. The first two 
correspond to a 2/1 resonance, the third to a 3/1 commensurability, while the 
latter is close to a 5/2 relationship. With regards to {\it Gliese 876}, 
numerical simulations (Laughlin and Chambers 2001, Lee and Peale 2002a) seem 
to indicate that these bodies are actually trapped in a stationary solution,
sometimes referred to as an {\it apsidal corotation}: they exhibit not only a 
libration of the resonant angle $\theta_1 = 2\lambda_2 - \lambda_1 - \varpi_1$,
but also an alignment of their major axes. System {\it HD 82943} also shows 
two planets in a 2/1 commensurability relation (Butler {\it et al.} 2002), 
although in this case it is unclear whether the observed motion is also an 
apsidal corotation or a simple $\theta_1$-libration. Similar doubts also exist 
for the planets located in other resonances.

Although to date there is no agreement as to the origin of these resonant 
configurations (see Perryman 2000 for a review of the different hypotheses), 
it is generally believed that the planets did not form at their present 
observed locations, but where driven (or trapped) into the resonances during a 
migration process triggered by some non-conservative force. Whether this 
orbital drift is still at work is also a matter of debate, although it is more 
plausible to assume that it stopped after the end of the planetary formation 
stage. However, and independently of how they reached this point, an important 
consequence of this type of configuration is that it constitutes a stabilizing 
mechanism for planetary orbits, especially if they have large eccentricities.

In the last few years, several studies have been performed to find initial 
conditions that yield stable stationary solutions. Since the observational data
usually does not yield explicit values for the masses, these tests are very 
important to estimate upper bounds for these parameters, as well as indications
for limits in the inclination of the orbital plane of the planetary system. 
One of the first analysis is due to Hadjidemetriou (2001), who studied the 
stability of numerically generated families of periodic orbits in the 
planar-elliptic 2/1 resonance. The results were later applied to both the 
{\it Gliese 876} and the {\it HD 82943} systems. Lee and Peale (2002a) 
presented a model for the {\it Gliese 876} system, and studied the capture of 
these planets into the resonance by an inward orbital migration. This 
trapping may explain the present apsidal corotation configuration, as well as 
the large eccentricity shown by both planets. The model was based on a 
Laplacian expansion of the disturbing function, truncated at third order in 
the eccentricities. The resulting equations were then used to discuss the 
existence and location of stationary solutions in the averaged (resonant) 
phase space. The authors concluded that two types of stable configurations 
exist: {\it (i)} for quasi-circular orbits such as the Io-Europa system, the 
corotation point lies in the axis $\varpi_1-\varpi_2=180$ degrees; {\it (ii)} 
for sufficiently large values of the eccentricities the stable solution occurs 
for $\varpi_1-\varpi_2=0$. In both cases, the resonant angle $\theta_1$ 
librates around zero. Similar results have also been presented by Murray 
{\it et al.} (2001).

It is important to bear in mind, however, that the Laplace expansion of the
disturbing function is not convergent for large values of the eccentricities.
In the case of the 2/1 resonance, the applicability of this model is restricted
to eccentricities below $0.17$ (see Ferraz-Mello 1994). For larger values, a 
truncated expansion may yield even qualitatively incorrect results 
(Beaug\'e 1994). Although this limit may not seem important in the case of our 
own Solar System where planets move in quasi-circular orbits, the same does 
not hold for extrasolar systems where highly elliptic motions abound. Even 
planets trapped in resonances show highly eccentric orbits. In 
{\it Gliese 876}, the best dynamical fits of observational data place the 
eccentricity of the inner planet at values of the order of $0.25 - 0.32$. 
For {\it HD 82943} it it seems even worse, since these values may reach 
$0.41 - 0.54$. 

The aim of the present work is to develop an adequate analytical model for the
study of planetary mean-motion resonances, even in the high-eccentricity case.
We will then search for all possible stable stationary solutions in terms of 
the orbital elements and masses, without mentioning possible resonance trapping
mechanisms. Thus, we shall not indulge in how the planets may have reached 
these solutions, our interest lying only in what these solutions look like and 
where they are located. The ultimate goal will be to present a ``chart'' or a 
``map'' of all the possible locations of planets in apsidal corotation, as a 
function of their eccentricities and mass ratios. With such a tool, we will be 
able to obtain a better idea of the permissible orbits of extrasolar planets. 
Under certain hypotheses, it will be possible to ascertain whether orbits 
deduced via data analysis of the observations are in fact plausible without 
having to resort to long-term numerical simulations for different masses.

This manuscript is divided as follows: In Section 2 we introduce the problem 
and the physical system under study. The analytical model is introduced in 
Section 3, where the conditions for corotational solutions are described. 
Application of the model to the 2/1 resonance, and results, are shown in 
Section 4. The case of the 3/1 resonance is discussed in Section 5. Finally, 
conclusions close the paper in Section 6.

\section{The General Three-Body Problem}

Suppose two planets of masses $m_1$ and $m_2$ orbiting a star of mass 
$M_0 \gg m_1,m_2$. Of course, the same dynamical system could be said to 
represent two massive satellites orbiting a planet. Both problems are 
equivalent from the dynamical point of view. However, since we are primarily 
interested in analyzing the case of extrasolar planets, throughout this paper 
we refer to $M_0$ as the star and $m_1,m_2$ as the planets.

This system is a particular case of the general three-body problem and is 
different of the restricted problems in which one of the bodies is supposed to 
have a negligible mass. We will consider two approximations. First, we will 
disregard oblateness effects of the central star and the planets, and thus 
consider only the mutual gravitational perturbations in the point-mass 
approximation. This limitation may be important if the secondary bodies orbit 
very close to the central body. Second, we will suppose all motion planar. 
This restriction does not seem to be significant, since the massive resonant
planetary satellites have practically no inclination with respect to the 
equator. For extrasolar planets, their mutual inclination is not known with 
any precision, and is usually taken equal to zero.

Let $a_i$ denote the semimajor axis of the $i^{\rm th}$-planet ($i=1,2$), 
$e_i$ the eccentricity, $\lambda_i$ the mean longitude and $\varpi_i$ the 
longitude of the pericenter. All orbital elements will be referred to the 
central star $M_0$. We will suppose $a_1 < a_2$, thus the subscript 2 will 
correspond to the outer orbiting body. This is a four degree of freedom system 
and its dynamical evolution can be specified by eight variables. In terms of 
the orbital elements of both planets, these can be given by the set 
$(a_1,a_2,e_1,e_2,\lambda_1,\lambda_2,\varpi_1,\varpi_2)$. Two integrals 
of motion exist: the total energy (or analogously, the Hamiltonian function
$F$), and the total angular momentum $J_{\rm tot}$.

Imagine now that both planets lie in the vicinity of a generic mean-motion 
commensurability $(p+q)/p$ with $q \neq 0$. In other words, let their mean
motions $n_i$ satisfy the relationship $(p+q)n_2 - pn_1 \sim 0$. In terms of 
their orbital elements, we can then define what is usually referred to as the 
{\it resonant} or {\it critical} angles of the system: 
\bea
\label{eq1}
q\sigma_1 &=& (p+q)\lambda_2 - p\lambda_1 - q\varpi_1 \\ 
q\sigma_2 &=& (p+q)\lambda_2 - p\lambda_1 - q\varpi_2 \nonumber
\eea
Since $\sigma_2-\sigma_1 = \varpi_1 - \varpi_2 = \Delta \varpi$, sometimes it 
is better to substitute the variable $\sigma_2$ with the difference in
longitude of pericenter $\Delta \varpi$. Similarly, since the resonant angle
$\sigma_1$ always appears in the disturbing function multiplied by the order
of the resonance $q$, it is also common practice to introduce new variables
$\theta_i = q \sigma_i$. With these changes in mind, we will consider 
$\theta_1$ and $\Delta \varpi$ as the slow angles of the problem. The fast 
angles, associated with the synodic period of the system, are usually 
eliminated by means of an averaging process.

\section{The Analytical Model}

In a recent study, Beaug\'e and Michtchenko (2002) presented an analytical
model for the Hamiltonian $F$ of the planetary three-body problem, At variance 
with classical expansions, this function is valid even for high values of the 
eccentricities of both planetary bodies, up to a limit of about 
$e_i \sim 0.5$. This limit varies according to the ratio $a_1/a_2$, being 
smaller for planets closer one to the other. We will then begin this section 
reviewing the form of this function.

\subsection{The Resonant Hamiltonian}

In Section 2, we mentioned that the averaged system in resonance can be 
specified by two angular variables, for example, $(\sigma_1,\sigma_2)$. Their 
canonical conjugates are given by:
\be
\label{eq2}
I_i = L_i \biggl(1 - \sqrt{1-e_i^2}\biggr) \hspace*{1.5cm} (i=1,2) .
\ee
Here $L_i = m'_i \sqrt{\mu_i a_i}$ is the modified Delaunay action related to 
the semimajor axis in Poincar\'e variables (see Laskar 1991), 
$\mu_i = G (M_0 + m_i)$, and $G$ is the gravitational constant. The factor 
$m'_i$ is a reduced mass of each body, given by:
\be
\label{eq3}
m'_i = {m_i M_0 \over m_i + M_0}  .
\ee
For small values of the eccentricities, 
$I_i \sim L_i e_i^2/2$. Thus, for fixed values of the semimajor axes, the 
momenta $I_i$ are proportional to the square of the eccentricities. 

In this resonant case, the Hamiltonian $F$ defines a two degree-of-freedom 
system in the canonical variables $(I_1,I_2,\sigma_1,\sigma_2)$. Apart from 
the Hamiltonian itself, two other integrals of motion exist in this averaged 
problem:
\bea
\label{eq4}
J_1 &=& L_1 + (p/q)   (I_1+I_2) = const \\
J_2 &=& L_2 - (1+p/q) (I_1+I_2) = const \nonumber 
\eea
such that $J_{\rm tot} = J_1 + J_2$ (Michtchenko and Ferraz-Mello 2001). With 
these definitions, the complete Hamiltonian of the system, averaged over 
short-period terms, can be expressed as:
\be
\label{eq5}
F = -\sum_{i=1}^2 {\mu_i^2 {m'}_i^3 \over 2 L_i^2} - F_1 ,
\ee
where the disturbing function $F_1$ can be expressed by means of the 
truncated series
\bea
\label{eq6}
F_1 &=& {G m_1 m_2 \over a_2} \sum_{j,k,l,u,s} R_{j,k,l,u,s} 
  (\alpha-\alpha_0)^j \cdot \\
  & \cdot& e_1^k e_2^l \cos{(uq \sigma_1 + s(\sigma_2 - \sigma_1))} \nonumber .
\eea
For a given resonance, the coefficients $R_{j,k,l,u,s}$ are constant for all 
initial conditions. The variable $\alpha = a_1/a_2$ denotes the ratio between 
semimajor axes, and $\alpha_0$ is its value at exact resonance. In the 
summations, the first four index vary from zero to a maximum value which is 
user-specified. The last index (i.e. $s$) takes both positive and negative 
values, whose limit can also be modified according to the desired precision. 
For details on the construction of this expression, the reader is referred to 
Beaug\'e and Michtchenko (2002).

\subsection{Stationary Solutions}

\begin{figure}[t]
\centerline{\includegraphics[width=20pc]{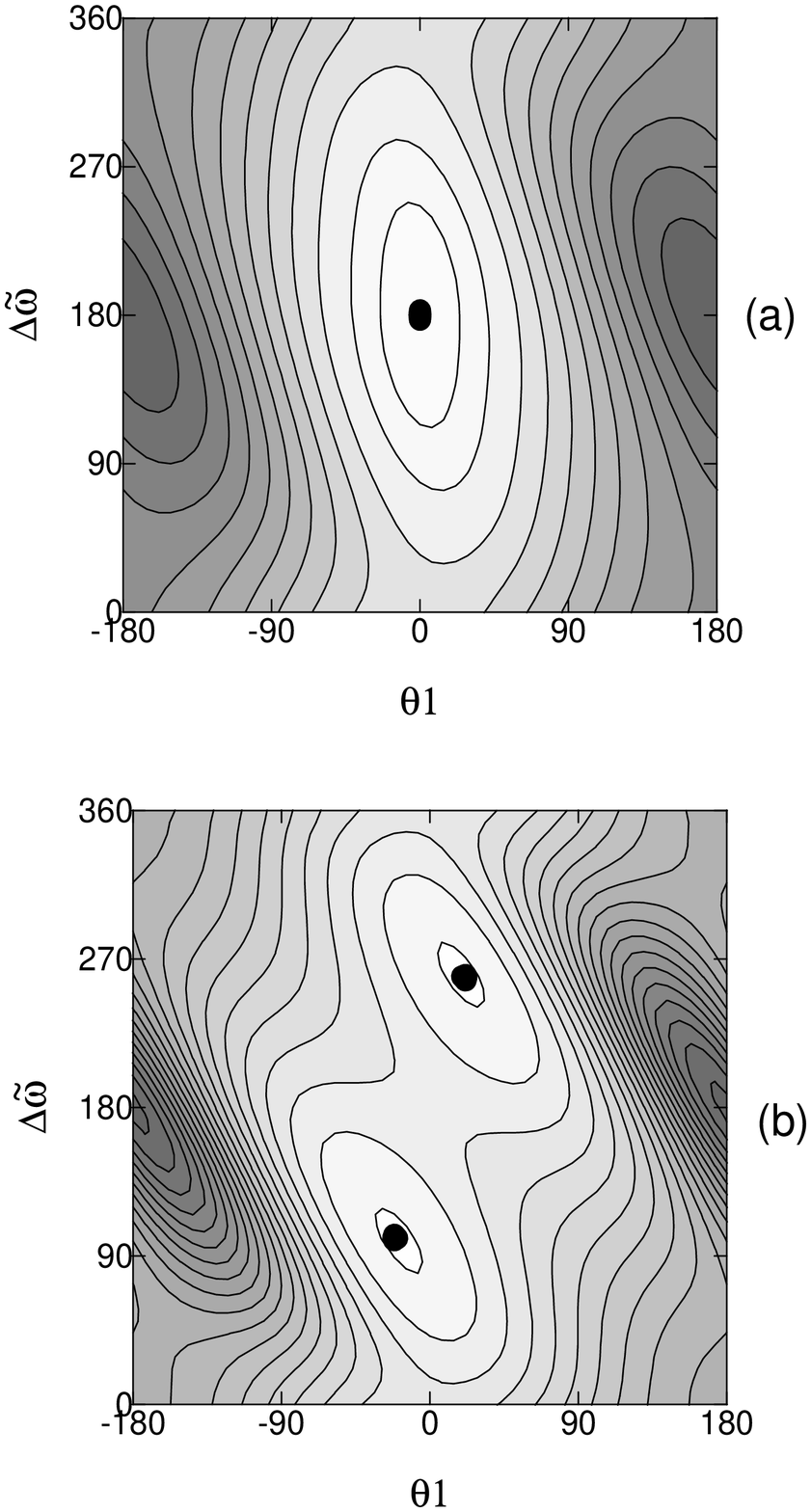}}
\caption{Level curves of the complete Hamiltonian function $F$, in the plane 
of $\Delta\varpi$ vs. $\theta_1$. Top graph corresponds to eccentricities 
equal to $e_1=e_2=0.02$, while bottom plot shows the results for $e_1=e_2=0.1$.
Lighter shades of gray indicate larger values of the function. Stable 
stationary solutions (maxima of $F$) are shown by dots.}
\label{fig1}
\end{figure}

Figure \ref{fig1} shows level curves of constant Hamiltonian $F$ for the 2/1
resonance, as function of both angular variables, and for fixed values of the 
planetary eccentricities. The top plot corresponds to $e_1=e_2=0.02$ while 
the bottom graph presents results for $e_1=e_2=0.1$. These contour plots give 
a general idea of the topology of the system, and help identify possible 
stationary solutions (seen here as full circles). This topology changes 
significantly from one plot to the other. The geometry of Figure \ref{fig1}(a) 
is what is expected in the so-called low-eccentricity regime, as is the case 
of the Io-Europa pair, and in which the stable stationary solution occurs at 
$\theta_1=\sigma_1=0$ and $\Delta \varpi = 180$ degrees. The bottom plot 
presents a different story. At the place of the previous stable solution, now 
appears an unstable solution (i.e. a saddle point). The solution bifurcated 
into two new ones. These new stable solutions are located outside the central 
axes of the figure and correspond, in the case shown, to the values  
$\theta_1 \simeq \pm 15$ degrees and $\Delta \varpi \simeq 180 \pm 80$ 
degrees. These are what we can call {\it asymmetric} stationary solutions. 

In order to study in detail the existence and location of all these stationary
solutions, we must make use of the equations of motion of the Hamiltonian
system. These are given by:
\be
\label{eq7}
{dI_i \over dt} = - {\partial F \over \partial \sigma_i} 
\hspace*{0.5cm} ; \hspace*{0.5cm} 
{d\sigma_i \over dt} =  {\partial F \over \partial I_i} 
\hspace*{1.0cm} (i=1,2) .
\ee
Thus, the stationary solutions will be given by the conditions:
\be
\label{eq8}
{\partial F \over \partial \sigma_i} = {\partial F \over \partial I_i} = 0
\hspace*{1.5cm} (i=1,2)
\ee
and their stability defined by the behavior of the Hessian of $F$ evaluated
at the point. These stationary solutions, or {\it apsidal corotation 
solutions}, are fixed points in the averaged system. When we reintroduce the 
short-period terms, they become periodic orbits in real space, whose period 
is simply the synodic period of the problem.

Before discussing these equations, let us simplify them introducing a change 
in units. First, we will consider $M_0$ as the unit of mass, and the planetary
masses $m_i$ will now be expressed in terms of the central mass. Second, we
will set $a_2$ as the unit distance. Thus, $a_1 = \alpha$. These
modifications will imply a new numerical value for $G$ but, otherwise, the
equations and coefficients are preserved. To search for the stationary 
solutions, let us discuss separately the derivatives with respect to the 
angles, and those with respect to the momenta. In the first one, since the 
two-body Hamiltonian contribution is independent of $\sigma_i$, we simply have 
that:
\bea
\label{eq9}
0 =  {\partial F_1 \over \partial \sigma_1} &=& G m_1 m_2 \sum_{j,k,l,u,s} 
       (uq-s) R_{j,k,l,u,s} (\alpha-\alpha_0)^j \nonumber \\
  &\cdot& e_1^k e_2^l \sin{(u\theta_1 + s\Delta \varpi )} \\
0 =  {\partial F_1 \over \partial \sigma_2} &=& G m_1 m_2 \sum_{j,k,l,u,s}  
        sR_{j,k,l,u,s} (\alpha-\alpha_0)^j \nonumber \\
  &\cdot& e_1^k e_2^l \sin{(u\theta_1 + s\Delta \varpi )} 
  \nonumber
\eea
where we have substituted $\theta_1 = q \sigma_1$ and 
$(\sigma_2-\sigma_1)=\Delta \varpi$. The right-hand 
sides of both equations are directly proportional to the product of the masses 
and the gravitational constant. Since the terms inside the sums are independent
of these parameters, the algebraic equations (\ref{eq9}) constrain the values 
of $\alpha-\alpha_0,e_1,e_2,\theta_1,\Delta\varpi$. If we forget, for the time 
being, the dependence with the semimajor axis (fixed at the exact resonance 
values, for instance), each given value of the pair $(e_1,e_2)$, lead to the 
values of $\theta_1$ and $\Delta\varpi$ such that equation (\ref{eq9}) is 
satisfied.

Independently of $(e_1,e_2)$, equations (\ref{eq9}) contain a set of trivial 
solutions, given by $\theta_1 = 0$ or $180$ and $\Delta\varpi = 0$ or $180$ 
(units are in degrees), although not all of them are stable. Depending on the 
eccentricities, in certain cases $\theta_1=0,\Delta\varpi=0$ will be stable. 
In others, the stability will appear in $\theta_1=0,\Delta\varpi=180$. In yet 
other cases, as seen in Figure \ref{fig1}(bottom), all those solutions will be 
unstable. 

If our expansion of the Hamiltonian function were restricted to a single 
harmonic in each angular variable (i.e. $u_{max} = s_{max} = 1$), as is 
usually the case in most simplified models, then symmetric solutions would be 
the only possible solution for system (\ref{eq9}). Nevertheless, if we 
consider a larger number of periodic terms, then it is possible, for certain 
values of $(e_1,e_2)$, to find zeros of these derivatives for values of 
$\theta_1$ and $\Delta\varpi$ different from $0$ or $180$ degrees.

Let us now concentrate of the remaining two equations of system (\ref{eq8}).
They are:
\bea
\label{eq10}
0 =  {\partial F \over \partial I_1} &=& 
     \sum_{i=1}^2 {\mu_i^2 {m'}_i^3 \over L_i^3} 
     {\partial L_i \over \partial I_1} + {\partial F_1 \over \partial I_1} 
 \nonumber \\
  &=&  {p \over q} \mu_1 n_1 - \biggl( 1 + {p \over q}\biggr) \mu_2 n_2
   + {\partial F_1 \over \partial I_1}  \\
0 =  {\partial F \over \partial I_2} &=& 
     \sum_{i=1}^2 {\mu_i^2 {m'}_i^3 \over L_i^3} 
      {\partial L_i \over \partial I_2} + {\partial F_1 \over \partial I_2} 
 \nonumber \\
  &=&  {p \over q} \mu_1 n_1 -  \biggl( 1 + {p \over q}\biggr) \mu_2 n_2
   + {\partial F_1 \over \partial I_2} \nonumber
\eea
where $n_i$ are the mean motions of the planets. Once again, the right-hand 
sides depend on both the orbital elements and the masses. We can write this 
explicitly saying that algebraic equations (\ref{eq10}) constrain the set 
\be
\label{eq11}
(e_1,e_2,\theta_1,\Delta\varpi,\alpha,m_1/m_2,m_1) .
\ee

In our own Solar System, the values of $m_1/m_2$ and $m_1$ are known with 
a high precision; they are not considered variables but known {\it parameters} 
of the problem. In such a case, it is common practice to choose a value of 
$\alpha$ and use equations (\ref{eq9})-(\ref{eq10}) to obtain the equilibrium 
values of $(e_1,e_2,\theta_1,\Delta\varpi)$ for each semimajor axis. The 
problem of extrasolar planets is different. Here, the individual masses of 
the bodies are usually not known, and only their ratio is estimated by the 
observational data. For this reason, a change of strategy is necessary. 

In the following sections we will determine the solutions as parametrized by
(supposedly unknown) values of $m_1$. In other words, we will consider a 
certain value for the mass of the inner body, solve the algebraic equations,
and then study the variation of these solutions for different values of this
parameter. In each case, we will analyze the plane defined by the 
eccentricities of both planets $(e_1,e_2)$, and for each point of this pair 
we will solve the algebraic conditions for apsidal corotations to obtain the 
(stable) stationary values of $(\theta_1,\Delta\varpi,\alpha,m_1/m_2)$. 
Consequently, for all the points in the plane of the orbital eccentricities, 
we will be able to plot level curves of the values of the angles and the ratio 
of the masses.

\section{Results for the 2/1 Resonance}

We begin analyzing the resonant solutions in which both longitudes of 
pericenter appear aligned or anti-aligned. Initially, we took the inner 
mass equal to $m_1=10^{-4} M_0$. Then, for each value of the pair $(e_1,e_2)$, 
we used equations (\ref{eq9})-(\ref{eq10}) to determine which symmetric 
apsidal solution corresponds to an extreme value of the Hamiltonian function. 
This procedure was repeated for a grid of eccentricities in the range 
$e_i \in [0,0.4]$.

\subsection{Apsidal Corotations with Anti-Aligned Pericenters}

\begin{figure}[t]
\centerline{\includegraphics[width=30pc]{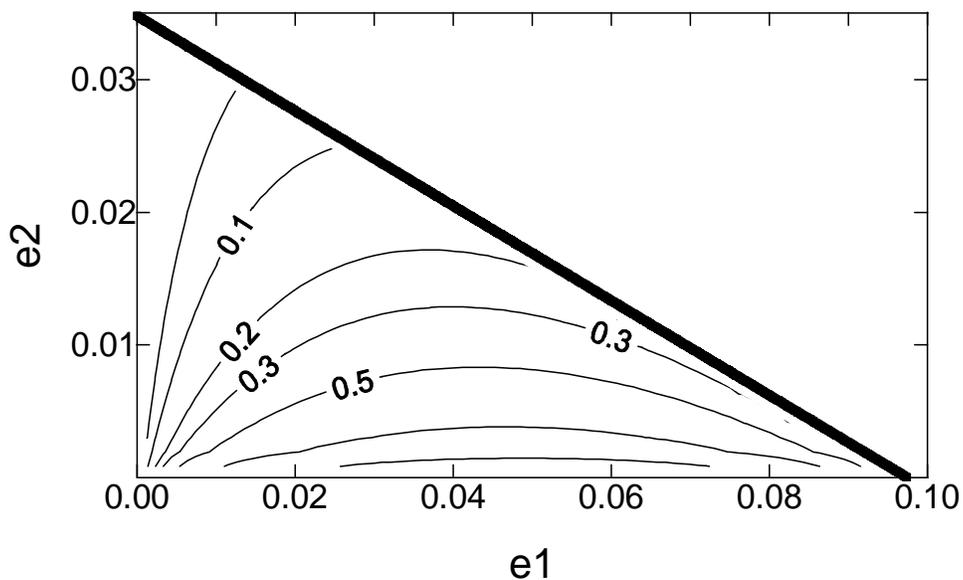}}
\caption{Anti-aligned corotation solutions ($\Delta \varpi=180$), for the 2/1 
resonance, in the plane $(e_1,e_2)$. The level curves correspond to constant 
values of the mass ratio $m_2/m_1$. The region is delimited by the thick black 
curve, above which no stable symmetric solution exist.}
\label{fig2}
\end{figure}

Results for anti-aligned apsidal corotations are shown in Figure \ref{fig2}. 
It can be seen that all solutions lie beneath a fairly straight broad curve, 
and are located at small values of the eccentricities. Inside this region, we 
have plotted the level curves of constant values of $m_2/m_1$. Thus, for 
example, for $e_1=0.06$, all solutions with $e_2 < 0.014$ correspond to stable 
apsidal corotations for some value of $m_2/m_1$. Conversely, if $e_2 > 0.014$, 
no stationary solution of this type exists, whatever the mass ratio of the 
planets.

The level curves themselves also yield important information. For example, they
show that in most of this region, the stable solutions correspond to cases 
where the mass of the exterior planet is smaller than that of the inner one. 
In the limit where $m_2/m_1 \rightarrow 0$, all solutions tend to $e_1=0$. In 
the opposite case, when $m_2 \gg m_1$, these corotations occur for small values
of $e_2$. Thus, in the limit where the general three-body problem tends towards
the restricted case, the eccentricity of the massive perturber must be very 
small to allow the pericenters to be anti-aligned.

\subsection{Apsidal Corotations with Aligned Pericenters}

\begin{figure}
\centerline{\includegraphics[width=30pc]{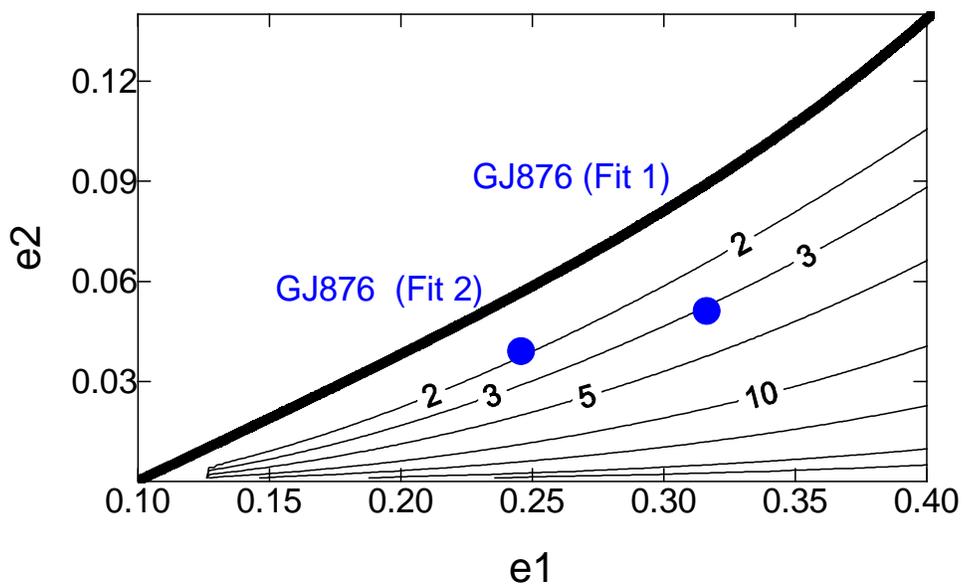}}
\caption{Same as previous figure, but for aligned apsidal corotations
(i.e. $\Delta \varpi=0$).}
\label{fig3}
\end{figure}

In Figure \ref{fig3} we present analogous results, now in the case where 
$\Delta\varpi=0$. Once again, all the solutions occur for values of $(e_1,e_2)$
beneath a broad black curve, but $e_1$ is no longer limited to small values. In
fact, this type of solution only occurs for eccentricities of the inner planet 
larger than $\sim 0.097$. The level curves of constant mass-ratio show that 
practically all the points correspond to $m_2 > m_1$. Solutions with 
$m_2 < m_1$ are also present, although these tend asymptotically to the thick
curve as the ratio $m_2/m_1$ approaches unity.

As a comparison of the model's results with numerical data, we also present 
in this graph (full circles) the best dynamical fits of the two planets of 
the {\it Gliese 876} system (Laughlin and Chambers 2001). What we call 
{\it Fit 1} refers to the dynamical fit with Keck+Lick observations and an 
inclination of the system's orbital plane equal to $\sin{i}=0.78$.  
{\it Fit2} corresponds to the Keck data only, and a value of $\sin{i}=0.55$. 
Both solutions fall well within the region of aligned pericenters, as predicted
by Lee and Peale (2002a). However, they do not correspond to the mass ratio 
corresponding to an exact apsidal corotation. From the observational data,
$m_2/m_1 \simeq 3.15$. Comparing it to our results, this shows a very good 
agreement with the level curves of {\it Fit 1}, although not so good for the 
second one. This discrepancy is probably due to the fact that the orbits 
deduced from the Keck data do not correspond to an exact corotation, but 
display a significant amplitude of libration of both angles. In other words, 
the solution does not lie at the maximum of $F$ (see Figure \ref{fig1}(top)) 
but corresponds to one of the closed curves around that point.

With regards to this point, it is important to recall the alleged origin of
this resonant configuration. If, as generally believed, these planets were 
trapped in the apsidal corotation through the action of non-conservative
forces, then it is expected that the bodies would have evolved towards an
extreme value of the total energy (i.e. Hamiltonian). In other words, the
observed corotation should correspond to a zero-amplitude solution. A look at 
the numerical simulations of Lee and Peale (2002a) shows that the adopted 
values of eccentricities and mass ratio of the planets does not yield such a 
zero-amplitude solution. This does not mean that the orbits are unstable; just 
that they are not extremes of the total energy. Consequently, either the 
dynamical fits are correct and the planets are not exactly at corotation, or 
the orbital elements of the planets must be modified to bring them to a 
stationary solution associated to the observed mass ratio.

Finally, from the practitioner's point of view, the most important aspect of 
Figures \ref{fig2} and \ref{fig3} is that they constitute a ``chart'' of 
possible solutions, as function of observable quantities such as the 
eccentricities and the ratio of masses. For example, if the observational data 
of a given planetary system in the 2/1 resonance yields a certain value of 
$e_1,e_2$ and $m_2/m_1$, we can check them against these plots and confirm 
whether they do indeed correspond to a stable apsidal corotation solution. 
This information can be obtained without having to resort to numerical 
simulations of the dynamical evolution of the system. 

\subsection{Asymmetric Stationary Solutions}

\begin{figure}[t]
\centerline{\includegraphics[width=23pc]{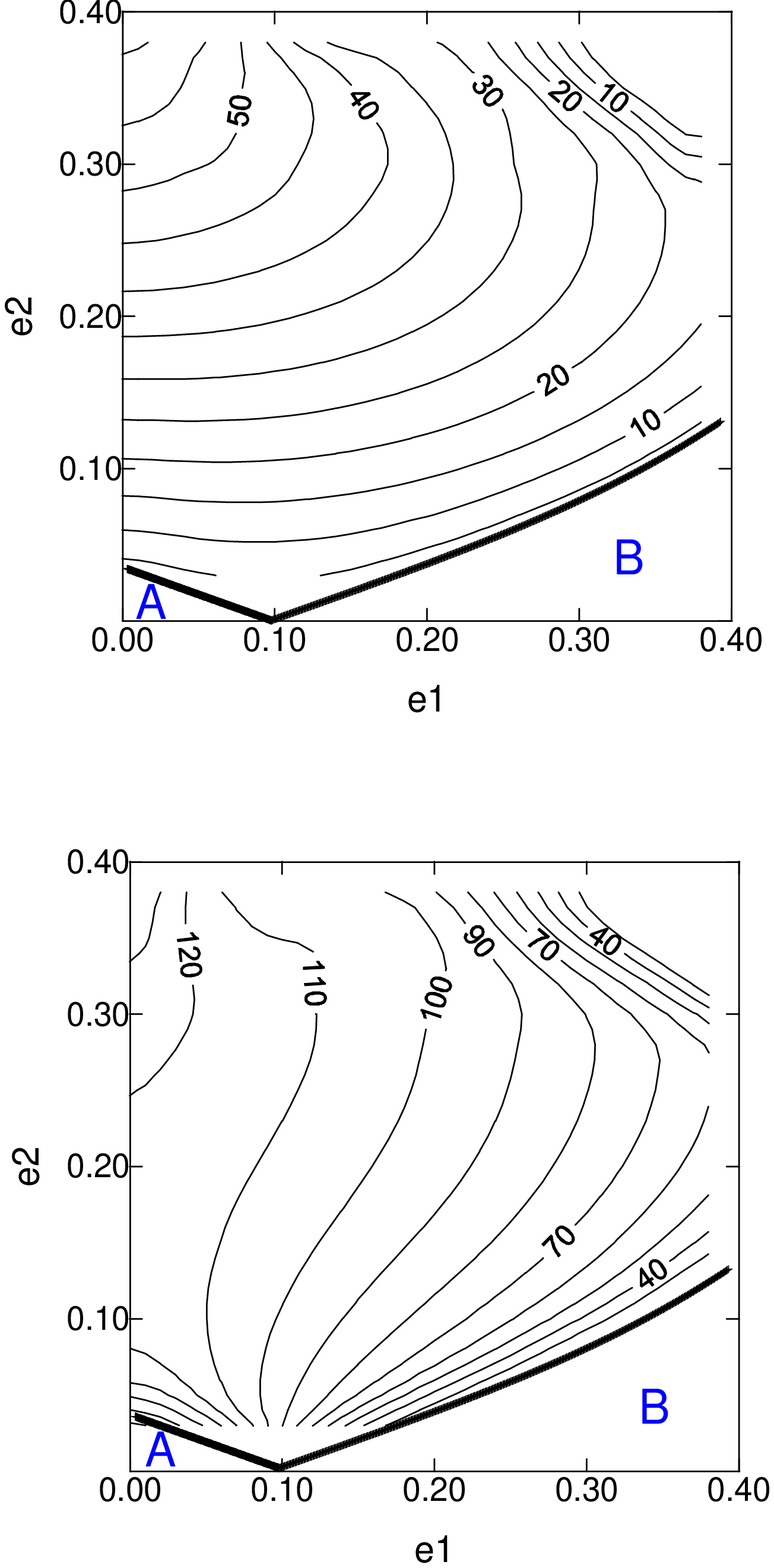}}
\caption{Asymmetric apsidal corotation solutions: Values of $\theta_1$ (top) 
and $\varpi_1-\varpi_2$ (bottom) in the plane $(e_1,e_2)$. Angles are given in 
degrees.}
\label{fig4}
\end{figure}

We have seen that the limit of the regions of solutions with anti-aligned and 
aligned pericenters are given by two curves stemming from the point 
$(e_1,e_2) \sim (0.097,0.000)$. The first are located in the interval 
$e_1 < 0.097$ and for $e_2$ beneath the corresponding curve. Aligned solutions 
are restricted to the open interval $e_1 > 0.097$. Above these broad curves 
all symmetric apsidal corotations are unstable, and we enter the realm of the 
asymmetric corotations. These new apsidal corotation solutions are shown in 
Figure \ref{fig4}, where once again we present solutions in the plane 
$(e_1,e_2)$. The top graph shows the values of $e_1,e_2$ corresponding to a 
same resonant angle $\theta_1$, while the bottom plot presents those with a 
same $\Delta\varpi$. All angles are given in degrees. 

\begin{figure}[t]
\centerline{\includegraphics[width=30pc]{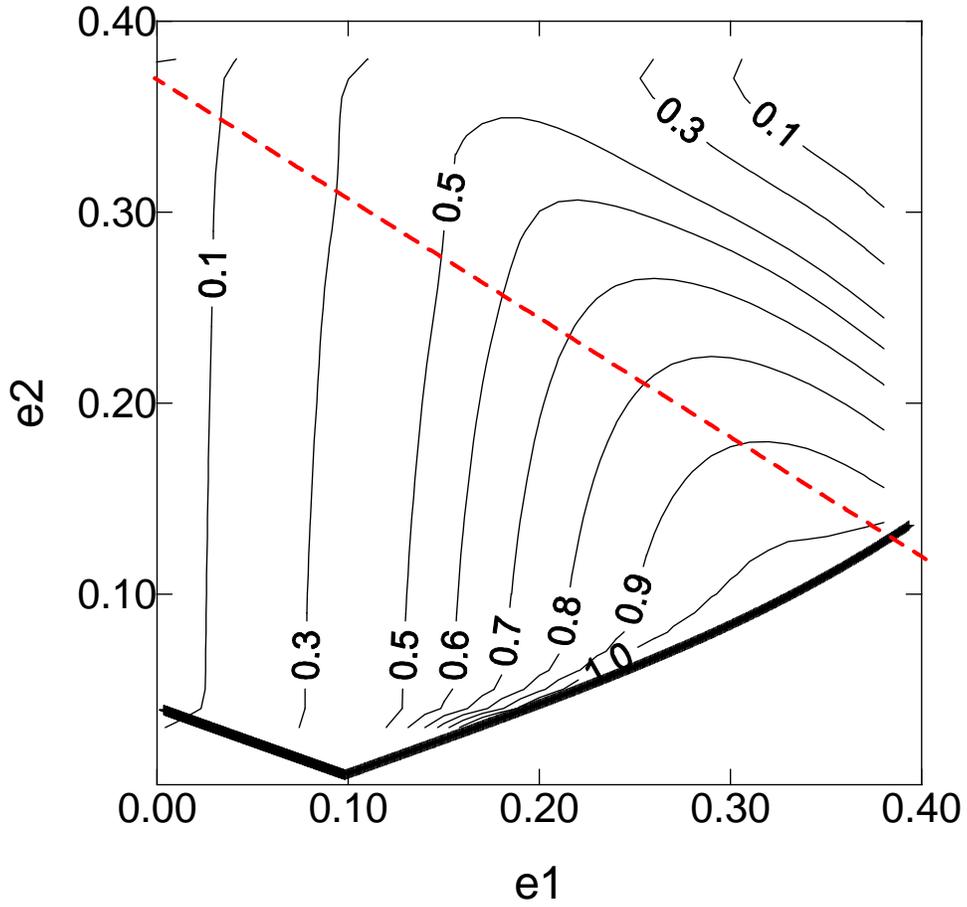}}
\caption{Level curves of the mass ratio $m_2/m_1$ in the asymmetric corotation
region. The diagonal broken line marks values of the eccentricities such that 
$a_1(1+e_1)=a_2(1-e_2)$.}
\label{fig5}
\end{figure}

These results were obtained solving the algebraic system (\ref{eq9}). We may 
use equations (\ref{eq10}) to assign a mass ratio $m_2/m_1$ to each pair of 
eccentricities. The results are shown in Figure \ref{fig5} once again in the 
form of a contour plot. We can see that practically all solutions occur for 
exterior planets smaller than interior ones. A small region of $m_2/m_1 > 1$ 
is present very close to the limit with the aligned apsidal corotations, 
although they exist only for values very close to unity. However, they are 
important in the sense that asymmetrical corotations can in fact exist in 
such a case. Also seen in this graph, the diagonal broken line shows the 
values of the eccentricities for which both orbits intersect, but are 
protected from a collision by the resonance.

Simultaneously with the mass ratio, equations (\ref{eq10}) also gives the 
stationary values of $\alpha$. With them, we may, now, return to the
algebraic system (\ref{eq9}) and determine corrected values of the angular
variables. These second-order corrections, however, never exceed $0.1$ degrees.
Thus, unless extremely precise values are needed, it is safe to consider all
the results in Figure \ref{fig4} as virtually independent of the
planetary masses. 

As a final note, let us point out that, due to the symmetries of the problem, 
if $\theta_1$ is a solution of the system, so is $360-\theta_1$. The same 
applies to the difference in pericenter. Although all these solutions may be 
grouped by plotting $\vert \theta_1 \vert=const$ and 
$\vert \Delta \varpi \vert=const$, it is important to stress that each solution
constitutes an individually distinct corotation created by the bifurcation in 
the location of the maximum of $F$ as shown in Figure \ref{fig1}. 

\subsection{Global View of the Apsidal Corotations}

These contour-plots give a global view of all possible solutions showing
apsidal corotations for eccentricities up to $0.4$. We can divide the plane in 
three regions. Symmetric solutions are categorized in regions {\bf A} and 
{\bf B}, according whether they correspond to an alignment or anti-alignment 
of the pericenters. The region above them corresponds to asymmetric 
corotations. With respect to the resonant angle, all symmetric solutions have
$\theta_1=0$. The non-symmetric solutions have  
$\theta_1 \sim 0$ for the values of $e_2$ immediately above the value 
corresponds to the broad curves. The stationary value of $\theta_1$ grows 
primarily with the eccentricity of the outer planet, reaching a maximum of 
about $60$ degrees for $(e_1,e_2) \simeq (0,0.4)$. The blank region at the 
upper right-hand side of the graph corresponds to high values of both 
eccentricities and is close to the limit of the model. The level curves of 
$\Delta \varpi$ show a different behavior. Since this angle is equal to $180$ 
degrees in the region {\bf A}, and equal to zero in {\bf B}, the asymmetric 
solutions also follow this trend and cover all the numerical values between 
these two extremes. At variance with the previous plot, the contours now seem 
to be more dependent on $e_1$ than on $e_2$.

\begin{figure}
\centerline{\includegraphics[width=20pc]{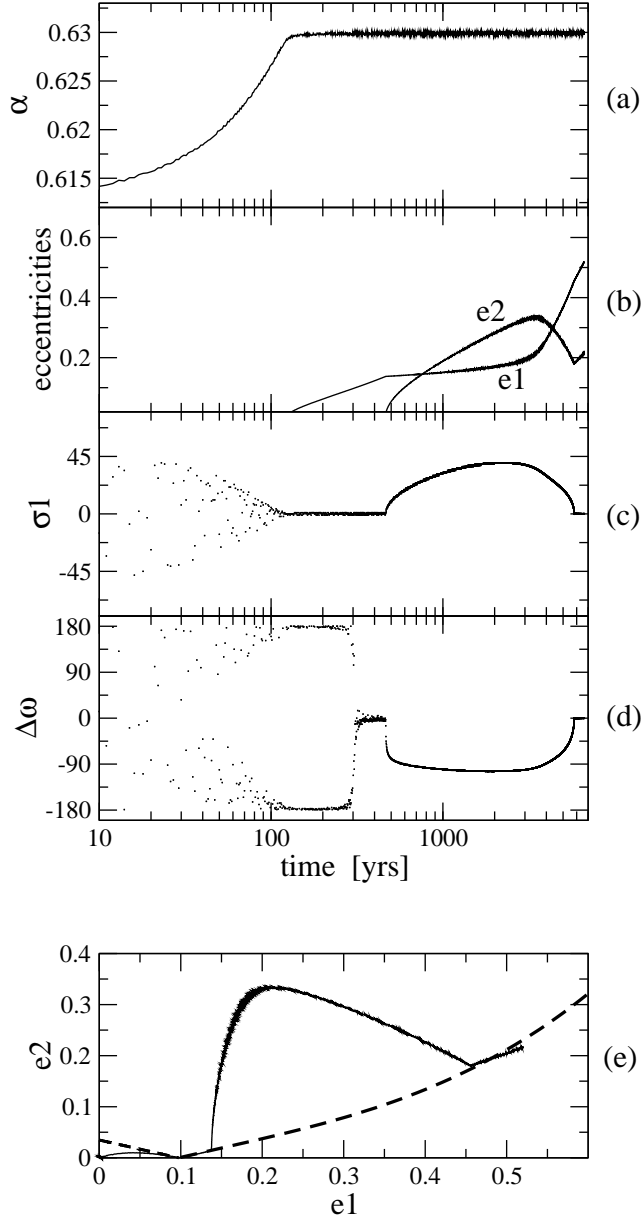}}
\caption{Numerical simulation of the capture process in the 2/1 resonance via 
an orbital migration of the inner planet. Mass ratio of both bodies is equal to
$m_2/m_1 \simeq 0.54$ (similar to the Io-Europa pair). {\bf (a)} Temporal 
evolution of the ratio of semimajor axes $\alpha=a_1/a_2$. {\bf (b)} 
Eccentricities of both planets as function of time. {\bf (c)} Resonant
angle $\theta_1$. {\bf (d)} Difference in longitude of pericenter 
$\Delta \varpi$. {\bf (e)} Plane of orbital eccentricities. Broad broken line 
shows separation between symmetric and asymmetric stationary solutions.}
\label{fig6}
\end{figure}

We have also performed several numerical simulations of the capture process in 
the 2/1 resonance, using different mass ratios and initial conditions 
(Ferraz-Mello {\it et al.} 2002). Figure \ref{fig6} shows a typical result 
obtained with $m_2/m_1 \simeq 0.54$, which is equal to the mass ratio of the 
Jovian satellites Io and Europa. On the top graph, we show the temporal 
evolution of $\alpha = a_1/a_2$. Both planets began the simulation in circular 
orbits with a mutual separation larger than that corresponding to the 2/1
commensurability. The semimajor axis of the inner body was then slowly 
increased due to the action of an anti-dissipative force pumping orbital energy
into the system. When the value of $\alpha$ reached the resonance (at 
approximately $t=100$) the capture occurred and, from this point onwards,
the ratio of semimajor axes remained fixed. Figure \ref{fig6}b shows the 
behavior of both eccentricities. When the trapping occurs, $e_1$ begins 
to grow while the eccentricity of the outer planet remains close to zero. A 
look at plots (c) and (d) show that this interval corresponds to symmetric 
apsidal corotations. The solution is first trapped in $\theta_1=0$ and 
$\Delta\varpi=180$ degrees. However, at $t \sim 300$ years, the relative
perihelia positions reverse and they become aligned ($\Delta \varpi=0$). This 
corresponds to an evolution from region {\bf A} to region {\bf B} (see 
Figure \ref{fig4}).

When the value of $e_1$ reaches $\sim 0.15$, the system leaves the domain of 
symmetric apsidal corotations and enters that of asymmetric ones. Both 
eccentricities now begin to grow significantly and the angular variables 
$\theta_1$ and $\Delta \varpi$ vary, taking values along the corresponding 
family of asymmetric apsidal corotations. This behavior continues up to near 
the end of the simulation, when the system returns to have anti-aligned apses.
Similar numerical results have been presented recently by 
Lee and Peale (2002b).

Finally, Figure \ref{fig6}(e) shows $e_2$ as function of $e_1$ for all the 
stationary solutions attained by the system during its evolution, allowing
a direct comparison with the analytical model (see Figure \ref{fig5}). As a 
reference, we have also plotted (broken lines) the limits between the domains
of symmetric and asymmetric solutions, as deduced from our equations.

\subsection{Dependence with the Actual Planetary Masses}

\begin{figure}
\centerline{\includegraphics[width=28pc]{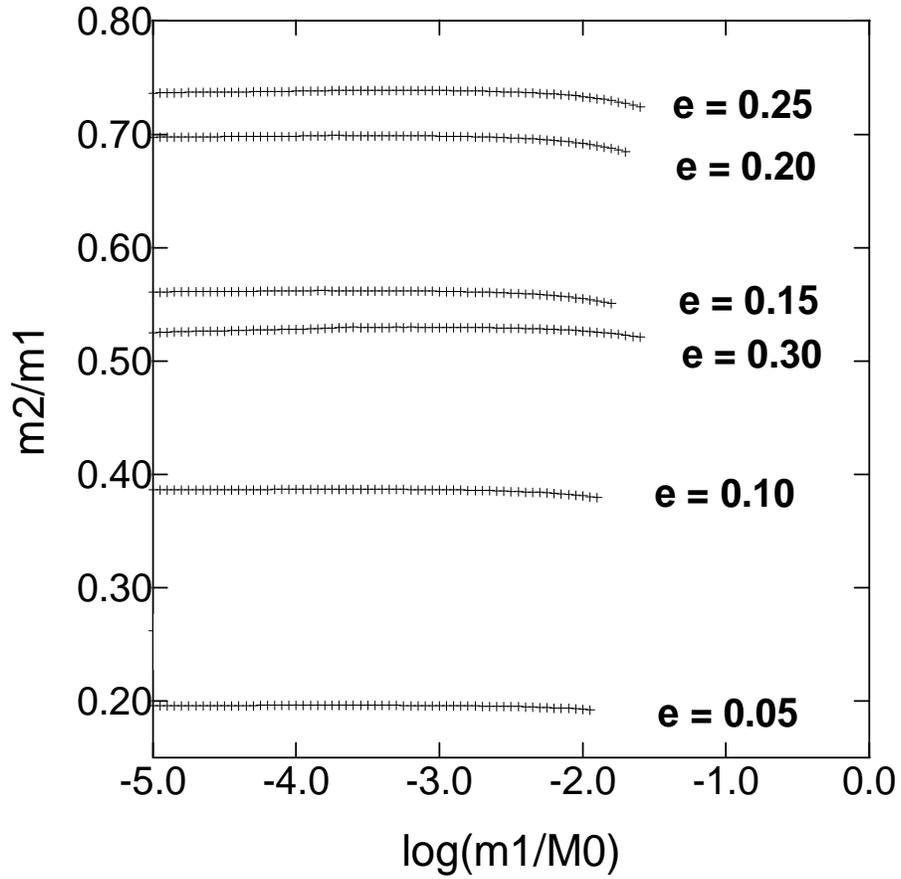}}
\caption{Variation of the mass ratio $m_2/m_1$ as function of $m_1/M_0$ in 
stationary solutions with $e_1=e_2=e$,}
\label{fig7}
\end{figure}

So far, we have determined all the solutions considering a fixed value of 
$m_1 = 10^{-4}$. We will now vary this mass and see its effect on our results. 
Recall that the stationary values of $\theta_1,\Delta \varpi$ are practically
invariant to this change. However, for a given pair $e_1,e_2$, the mass ratio 
$m_2/m_1$ may depend on this parameter. To see this dependence, we analyzed a 
few corotational solutions (chosen such that $e_1=e_2$). For each, we took 
values of the inner mass in the range $\log{(m_1/M_0)} \in [-5,0]$, and 
determined the mass ratio corresponding to a stationary solution. The results 
are shown in Figure \ref{fig7}. We can see that the dependence of the solution 
with $m_1$ is very weak, never exceeding $1\%$. In other words, the 
corotational solutions determined for one value of $m_1$ will suffer 
practically no change if this value is changed.

This has two main consequences. First, the results presented in Figures 
\ref{fig2}-\ref{fig4} are practically independent of the individual masses 
of the system. Thus, they are very general and should be applicable to 
practically any planetary system found in a 2/1 resonance. However, this 
lack of sensitivity of the results implies also that it is not possible to 
deduce the individual masses from the fact that a given system lies in an
exact apsidal corotation.

\begin{figure}
\centerline{\includegraphics[width=30pc]{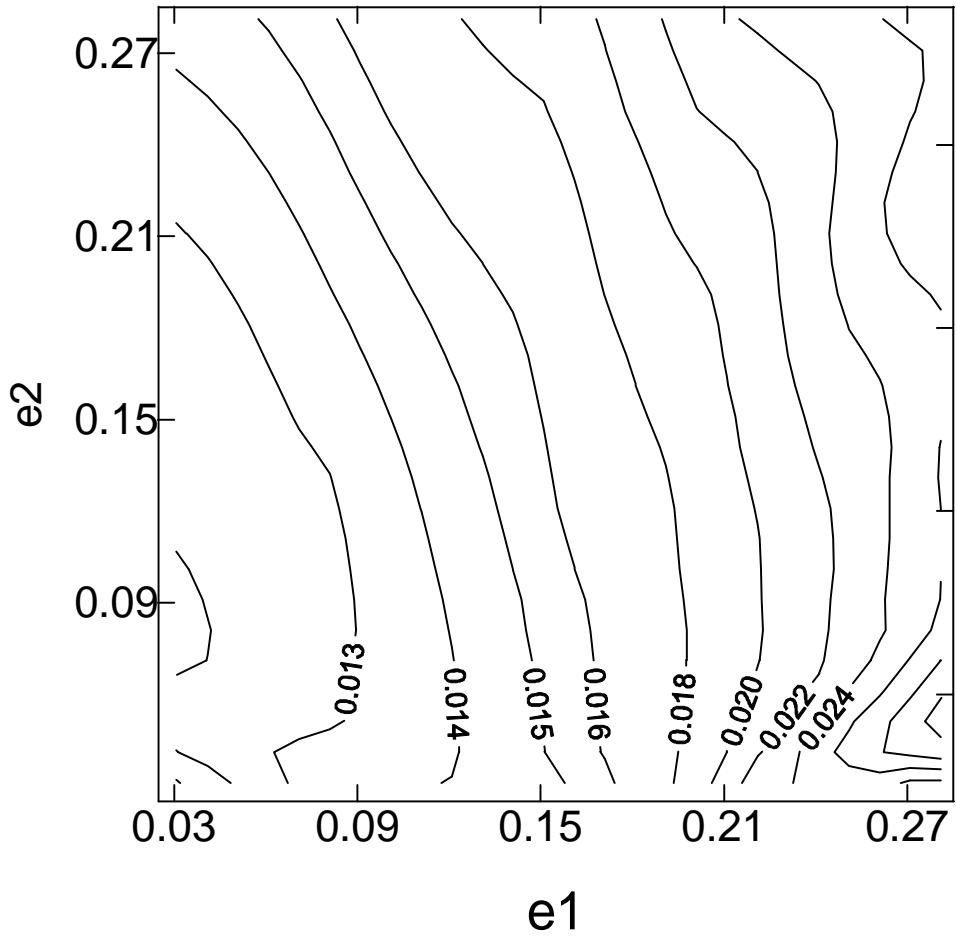}}
\caption{In the plane $(e_1,e_2)$, maximum value of $m_1/M_0$ for which stable 
corotations exist.}
\label{fig8}
\end{figure}

Second, note that for each value of $e$ there exists a maximum value of $m_1$ 
for which the stable corotations exist. The variation of this critical value 
of $m_1$ in the plane $(e_1,e_2)$ is shown in Figure \ref{fig8} as level 
curves. In the interval of eccentricities considered, all limit values are 
very similar, and of the order of $\sim 0.01 - 0.02 M_0$. If the individual 
mass of the inner planet is larger than this limit, then the apsidal corotation
solutions become unstable. This places some dynamical limit on the values of 
the individual planets showing apsidal corotations, and not just on their mass 
ratio. Although not very restrictive, this limit is enough to exclude for 
brown dwarfs.

\section{The 3/1 Resonance}

Extrasolar planets are not only found in the vicinity of the 2/1 
commensurability, but also around other resonances. In particular, the 
{\it 55 Cnc} system seems to contain three planets, two of which orbiting very 
close to a 3/1 resonance. Different four-body dynamical fits yield 
eccentricities of the order of $0.013 - 0.039$ for the inner mass, and 
$0.080 - 0.157$ for the outer. Masses are of the order of $0.83 M_{Jup}$ and 
$0.20 M_{Jup}$, respectively.

In this section we will determine the apsidal corotations of the 3/1
commensurability, in the same manner as discussed for the 2/1 resonance. The 
equations of motion and conditions for equilibrium are analogous; the only
care we must take is to change the values of $p$ and $q$. For this 
configuration, $p=1$ and $q=2$. It is important to recall that, in the 
asteroidal case, all exterior resonances of the type $(p+q)/p= j$ (with $j$
positive integer) have asymmetric librations (Morbidelli {\it et al.} 1995).

\begin{figure}[t]
\centerline{\includegraphics[width=25pc]{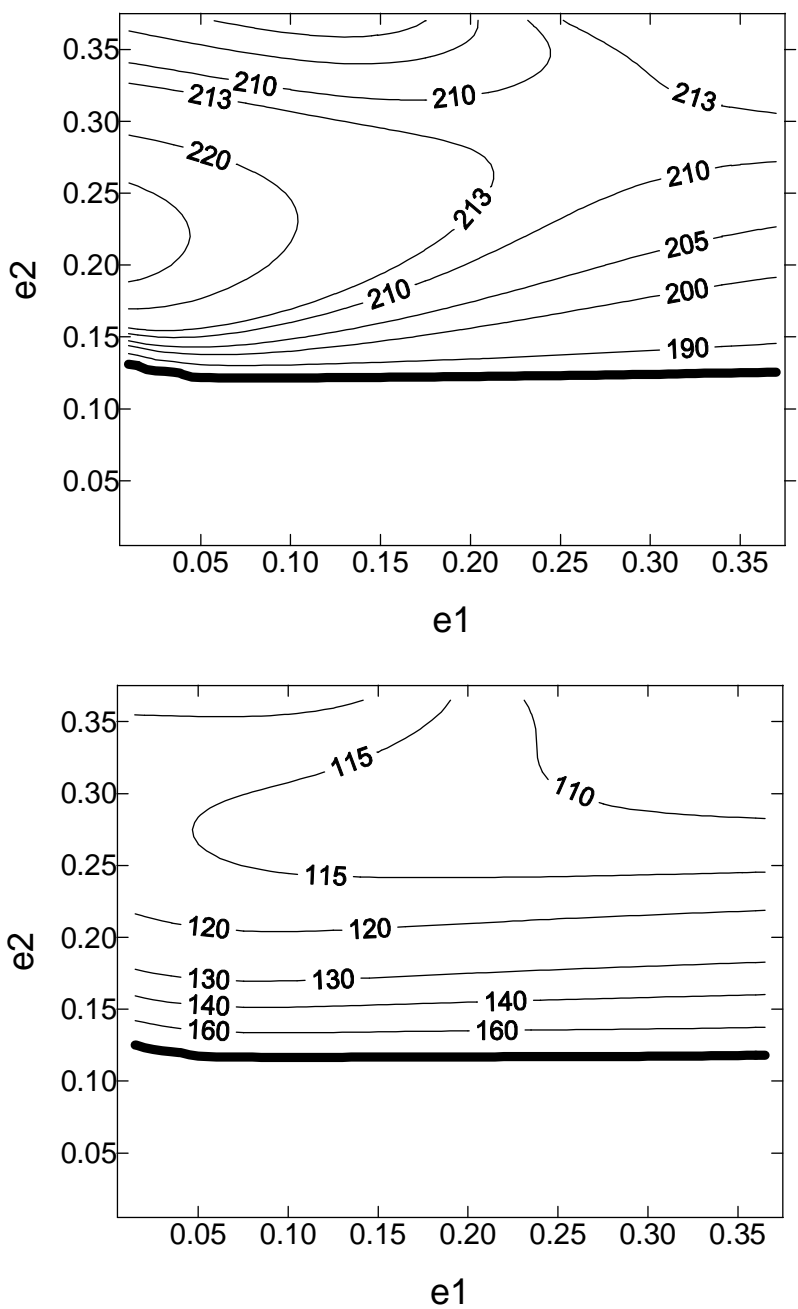}}
\caption{Level curves of $\theta_1$ (top) and $\Delta \varpi$ (bottom) in the 
eccentricities-plane for the 3/1 mean-motion resonance. Separation between
symmetric and asymmetric regions is given by the broad black curves, below 
which the stationary solutions correspond to $\theta_1=180$ and 
$\Delta \varpi = 180$ degrees.}
\label{fig9}
\end{figure}

Figure \ref{fig9} shows all the stable stationary values of the angular 
variables $\theta_1$ and $\Delta \varpi$ in the plane $e_1,e_2$. Two 
independent regions are evident. For all values of $e_2 <\,\sim 0.13$, apsidal 
corotation solutions are symmetric and occur for $\theta_1 = 180$ and 
$\Delta \varpi = 180$ degrees. However, if the eccentricity of the inner 
planet is larger than the limit $e_2 \simeq 0.13$, the stationary solutions 
become asymmetric. It is interesting to note that, at variance with the 2/1 
case, the limit between both domains is practically independent of $e_1$.

The mass ratio of both planets corresponding to these solutions is shown in 
Figure \ref{fig10}. The contour plots are close to piecewise straight lines, 
with a slope discontinuity at the boundary of the two regions. One may note 
that $e_2 \rightarrow 0$, the value of $m_2/m_1$-ratio of the apsidal
corotation solutions approaches and the used model becomes the so-called
interior asteroid problem. Conversely, when $e_1 \rightarrow 0$ the mass ratio 
is smaller than unity. Although symmetric solutions show a preponderance of 
systems with $m_2 > m_1$, and the inverse occurs in asymmetric orbits, in both 
cases we can see examples with all values of the mass ratio.

\section{Conclusions}

We have presented, in this paper, analytical and numerical results on the 
existence and location of stable stationary solutions, in the averaged 
planetary three-body problem, around the 2/1 and 3/1 mean-motion resonances.
We have shown the existence of two distinct type of solutions. The first of
these are symmetric apsidal corotations with , with $\varpi_1 - \varpi_2 = 0$ 
(aligned periapses) or $\varpi_1 - \varpi_2 = 180$ degrees (anti-aligned 
periapses); In  the 2/1 resonance, $\theta_1 = 0$, while in the 3/1 resonance 
$\theta_1 = 180$ degrees. These solutions are well known from classical 
studies dating from the works of Laplace in the nineteenth century. A second 
type of equilibrium solutions has also been found, this time characterized by 
asymmetric configurations, in which both the resonant angle $\theta_1$ and 
$\varpi_1 - \varpi_2$ librate around values which are not 0 or 180 degrees. 
Although asymmetric resonant solutions are known to occur in exterior 
resonances of the restricted three-body problem (see Beaug\'e 1994), only 
recently has there been any indication that they may also be present in the 
planetary case (Ferraz-Mello, 2002; Lee and Peale 2002b).

\begin{figure}[t]
\centerline{\includegraphics[width=30pc]{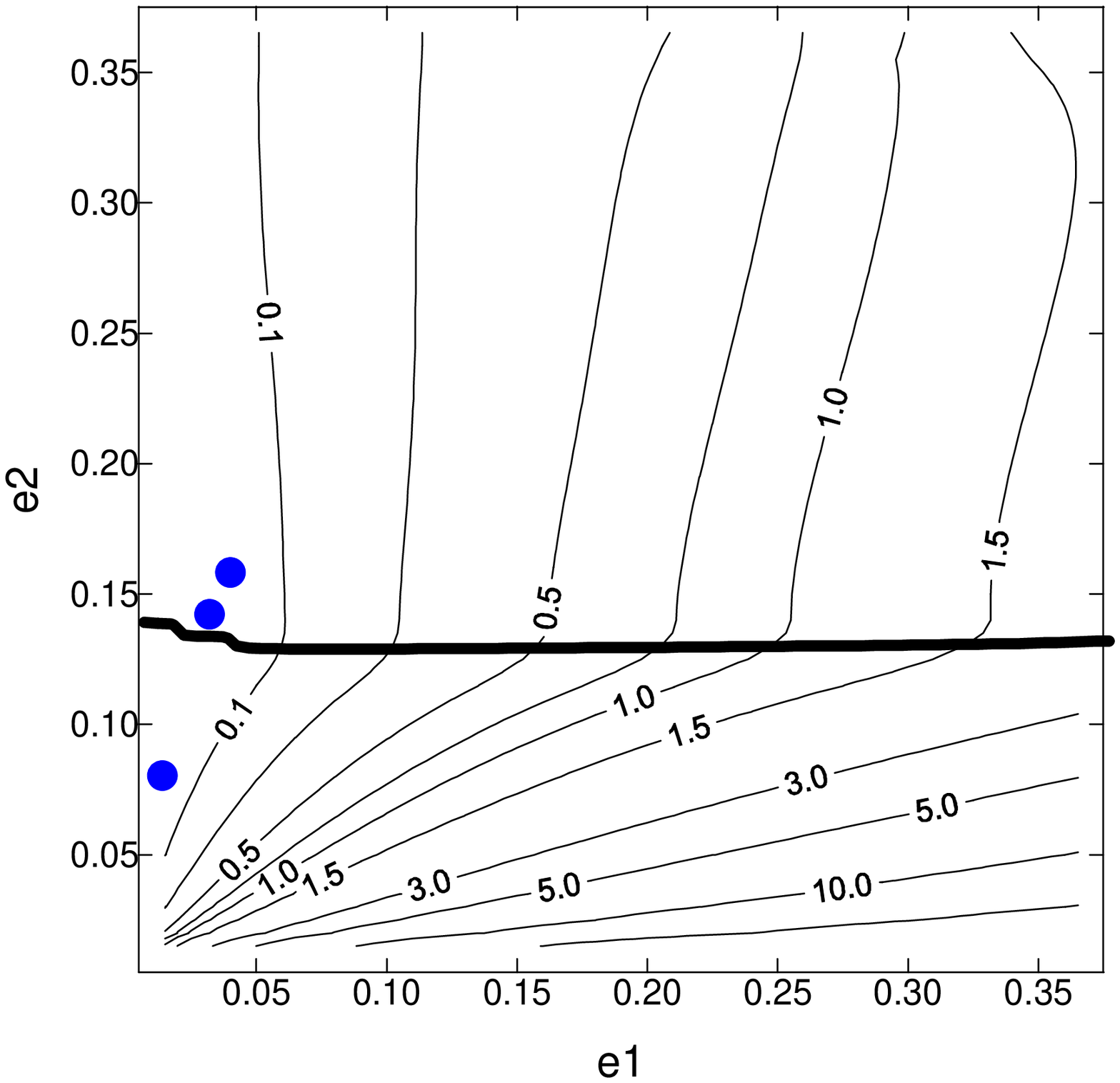}}
\caption{Mass-ratio $m_2/m_1$ of the stationary solutions in the $(e_1,e_2)$
plane. Note the change in behavior between the apsidal region (below the 
broad curve) and the asymmetric zone (above). In full circles we also show 
the three best dynamical fits of the {\it 55Cnc} system.}
\label{fig10}
\end{figure}

The plots we have presented for both commensurabilities are useful tools for 
the research in extrasolar planets. On one hand, they allow for a simple 
method for testing whether a certain dynamical fit of observational data is 
consistent with stable corotations. This information is obtained without 
having to resort to any numerical simulation of the equations of motion. As 
an additional example of this application, in Figure \ref{fig10} we have 
plotted (in full circles) the three best dynamical fits of the two resonant 
planets of the {\it 55Cnc} system (Marcy {\it et al.} 2002). Only one of them 
corresponds to a symmetric apsidal corotation, while the other two lie in 
the asymmetric domain. Furthermore, all of them correspond to mass ratios of 
the order of $0.05$, although from observational data it is estimated that 
$m_2/m_1 \sim 0.24$. 

It is conceivable that, although these planets may be trapped in the 3/1 
resonance, they do not lie in an apsidal corotation solution but in a 
$\theta_1$-libration only. In other words, they could exhibit a libration of 
the resonant angle $\theta_1$ but a circulation of $\theta_2$ and, thus, a
circulation of the relative perihelion positions. The problem with this 
hypothesis is that all numerical simulations of the capture process of 
extrasolar planets have resulted in apsidal corotations and there has been 
no reported case of capture in $\theta_1$-libration only. This does not mean 
such solutions do not exist; perhaps their capture probability is small, or 
they are very sensitive to the modelization of the non-conservative force that 
drives the orbital migration. Hopefully, further simulations will shed new
light on this question.

Finally, we have seen that although the solutions are only weakly dependent on 
the individual masses of the planets, there does exist a maximum value of $m_i$
for which these are stable. This limit allows us to estimate upper bounds for
the values of the masses, and thus give qualitative information about the
maximum inclination of the orbital plane of the planetary systems.

\acknowledgments{
This work has been supported by the Brazilian National Research Council -CNPq-
through the fellowships 300953/01-1 and 300946/96-1, as well as the S\~ao Paulo
State Science Foundation FAPESP. The authors gratefully acknowledge the support
of the Computation Center of the University of S\~ao Paulo (LCCA-USP) for the 
use of their facilities.}

\newpage

\begin{thebibliography}{}

\bibitem[]{}
Beaug\'e, C. (1994). Asymmetric librations in exterior resonances. 
{\it Cel. Mech. \& Dynam. Astron.}, {\bf 60}, 225-248.

\bibitem[]{}
Beaug\'e, C. and Michtchenko, T.A. (2002). Modeling the high-eccentricity 
planetary three-body problem. {\it MNRAS }, submitted.

\bibitem[]{}
Butler, P., Marcy, G., Vogt, S., Tinney, C., Jones, H., McCarthy, C., Penny, 
A., Apps, K. and Carter, B. (2002). On the double planet system around 
HD 83443. {\it ApJ}, {\bf 578}, 565-572.

\bibitem[]{}
Ferraz-Mello, S. (1994). The convergence domain of the Laplacian expansion of 
the disturbing function. {\it Cel. Mech. \& Dynam. Astron.}, {\bf 58}, 37-52.

\bibitem[]{}
Ferraz-Mello, S. (2002). Tidal Acceleration, Rotation and Apses Alignment in 
Resonant Extra-Solar Planetary Systems {\it Bull. A.A.S. }, in press. 

\bibitem[]{}
Ferraz-Mello, S., Michtchenko, T.A. and Beaug\'e, C. (2003). Dynamics of the
convergent migration of two planets under the action of anti-dissipative
forces. {\it Cel. Mech. \& Dynam. Astron.}, {\bf 87} to be published.

\bibitem[]{}
Hadjidemetriou, J. (2001). Resonant periodic motion and the stability of 
extrasolar planetary systems. {\it Cel. Mech. \& Dynam. Astron.}, {\bf 83}, 
141-154.

\bibitem[]{}
Laskar, J. (1991). Analytical framework in Poincar\'e variables for the motion 
of the Solar System. In: {\it Predictability, Stability and Chaos in N-Body 
Dynamical Systems} (A. Roy, Ed.). NATO Asi Series, Vol. B272, pp. 93-112,  
Plenum Press, New York.

\bibitem[]{}
Laughlin, G. and Chambers, J. (2001). Short-term dynamical interactions among 
extrasolar planets. {\it ApJ}, {\bf 551}, L109-L113.

\bibitem[]{}
Lee, M.H. and Peale, S.J. (2002a). Dynamics and origin of the 2:1 orbital 
resonances of the GJ 876 planets. {\it ApJ}, {\bf 567}, 596-609.

\bibitem[]{}
Lee, M.H. and Peale, S.J. (2002b). Extrasolar planets and mean-motion 
resonances. In Scientific Frontiers in Research on Extrasolar Planets.
Eds. D. Deming and S. Seager, (ASP Conf. Series), in press.

\bibitem[]{}
Marcy, G.W., Butler, R.P., Fischer, D., Laughlin, G., Vogt, S.S., Henry, G.W. 
and Pourbaix, D. (2002). A planet at 5 AU around 55 Canci. {\it ApJ}, in 
press.

\bibitem[]{}
Michtchenko, T.A. and Ferraz-Mello, S. (2001). Modeling the 5:2 mean-motion
resonance in the Jupiter-Saturn planetary system. {\it Icarus}, {\bf 149},
357-374.

\bibitem[]{}
Morbidelli, A., Thomas, F. and Moons, M. (1995). The resonant structure of the 
Kuiper belt and the dynamics of the first five trans-neptunian objects. 
{\it Icarus}, {\bf 118}, 322-340.

\bibitem[]{}
Murray, N., Paskowitz, M. and Holman, M. (2001). Eccentricity evolution of 
resonant migrating planets. {\it ApJ}, {\bf 565}, 608-620.

\bibitem[]{}
Perryman, M.(2000). Extra-solar planets. {\it Rep. on Progress in Phys.}, 
{\bf 63}, 1209.

\end{thebibliography}
\end{document}